\documentclass[letterpaper,titlepage,11pt]{article}
\usepackage{hyperref}

\usepackage{amssymb,amsmath,amsfonts}
\usepackage{epsfig}

\def\clock{{\count0=\time
           \divide\count0 60
           \ifnum\count0<10 0\fi\the\count0
           \multiply\count0 -60 \advance\count0 \time
           :\ifnum\count0<10 0\fi \the\count0
         }}
\newcommand{\timestamp}{{\small\vbox{\hbox{\tt\jobname.tex}
\hbox{\the\day/\the\month/\the\year, \clock}}}}

\newcommand{\be}{\begin{equation}} \newcommand{\ee}{\end{equation}}
\newcommand{\bea}{\begin{eqnarray}} \newcommand{\eea}{\end{eqnarray}}

\newcommand{\id}{\hbox{1\kern-.27em l}}
\newcommand{\sid}{\hbox{\scriptsize1\kern-.27em l}}

\newcommand{\we}{\kern-.1em\wedge\kern-.1em}
\newcommand{\scal}{\kern-.13em\cdot\kern-.13em}

\newcommand{\II}{I\kern-.09em I}

 \newcommand{\de}{\delta}

\newcommand{\R}{\mathbb{R}}

\newcommand{\pr}{\partial_r}
\newcommand{\pz}{\partial_z}

\newcommand{\spa}{\ , \ \ }

\numberwithin{equation}{section}

\begin{document}

\begin{titlepage}

\rightline{\vbox{\small\hbox{\tt hep-th/0309230} }}
\vskip 3cm

\centerline{\Large \bf Phase Structure of}
\vskip 0.1cm
\centerline{\Large \bf Black Holes and Strings on Cylinders}

\vskip 1.6cm
\centerline{\bf Troels Harmark and Niels A. Obers}
\vskip 0.5cm
\centerline{\sl The Niels Bohr Institute}
\centerline{\sl Blegdamsvej 17, 2100 Copenhagen \O, Denmark}

\vskip 0.5cm

\centerline{\small\tt harmark@nbi.dk, obers@nbi.dk}

\vskip 1.6cm

\centerline{\bf Abstract}
\vskip 0.2cm
\noindent
We use the $(M,n)$ phase diagram recently introduced in hep-th/0309116 to
investigate the phase structure of black holes and strings on
cylinders.
We first prove that any static neutral black object on a cylinder can be
put into an ansatz for the metric
originally proposed in hep-th/0204047, generalizing a result of
Wiseman. Using the ansatz, we then show that all branches of
solutions obey the first law of thermodynamics and that
any solution has an infinite number of copies. The consequences
of these two results are analyzed.
Based on the new insights and the known branches of solutions,
we finally present an extensive
discussion of the possible scenarios for the Gregory-Laflamme
instability and the black hole/string transition.


\end{titlepage}

\pagestyle{empty}

\tableofcontents

\newpage

\pagestyle{plain}
\setcounter{page}{1}

\section{Introduction}

In \cite{Harmark:2003dg} a new type of phase diagram
for neutral and static black holes and black strings on cylinders
$\R^{d-1} \times S^1$ was introduced.%
\footnote{Recently, the paper \cite{Kol:2003if} appeared which
contains similar considerations.}
In this paper we continue to examine various aspects of the
phase structure of black holes and strings on cylinders
using this phase diagram. We furthermore establish new
results that deepen our understanding of the phase structure.

The phase structure of neutral and static black holes and black
strings on cylinders $\R^{d-1} \times S^1$ has proven to be very
rich, but is also largely unknown. We can categorize the possible
phase transitions into four types:
\begin{itemize}
\item Black string $\rightarrow$ black hole
\item Black hole  $\rightarrow$ black string
\item Black string  $\rightarrow$ black string
\item Black hole  $\rightarrow$ black hole
\end{itemize}
Since the event horizons of the black string and black hole
have topologies $S^{d-2} \times S^1$ and $S^{d-1}$, respectively,
we see that the first two types of phase transitions
involve changing the topology of the horizon, while the last two
do not.

Gregory and Laflamme \cite{Gregory:1993vy,Gregory:1994bj}
discovered that light uniform black strings are classically
unstable. In other words
strings that are symmetric around the cylinder, are unstable
to linear perturbations when the mass of the string
is below a certain critical mass.
This was interpreted to mean that a light uniform string decays
to a black hole on a cylinder since that has higher entropy,
thus being a transition of the first type above.

However, Horowitz and Maeda argued in \cite{Horowitz:2001cz}
that the topology of the event horizon cannot change through
a classical evolution. This means that the first two types of
transitions above cannot happen through classical evolution.
Therefore, Horowitz and Maeda conjectured
that the decay of a light uniform string
has an intermediate step: The light uniform
string decays to a non-uniform string, which then eventually decays
to a black hole.%
\footnote{Note that the classical decay
of the Gregory-Laflamme instability was studied numerically
in \cite{Choptuik:2003qd}.}
Thus, we first have a transition of the third type, then afterwards
a transition from non-uniform black string to a black hole which
is of the first type.

The conjecture of Horowitz and Maeda in \cite{Horowitz:2001cz}
prompted a search for non-uniform black strings, and a branch of
non-uniform string solutions was found in
\cite{Gubser:2001ac,Wiseman:2002zc}.%
\footnote{This branch of non-uniform string solutions was
in fact already discovered by Gregory and Laflamme
in \cite{Gregory:1988nb}.}
However, this new branch of non-uniform string solutions
was discovered to have lower entropy than uniform strings
of the same mass, thus a given non-uniform string will
decay to a uniform string (i.e. a process of the third type
above).

If we consider black holes on a cylinder, it is clear that
they cannot exist beyond a certain critical mass, since at some point
the event horizon  becomes too large to fit on
a cylinder. Therefore, there should be a transition
from large black holes to black strings, i.e. a transition
of the second type above. However, it is not known how this transition
precisely works.

Finally, as an example of a transition of the fourth type above,
one can consider two black holes on a cylinder which are put opposite
to each other. This configuration exists as a classical solution but
it is unstable. Thus,
the two black holes will decay to one black hole.

Several suggestions for the phase structure of
the Gregory-Laflamme instability, the black hole/string transitions
and the uniform/non-uniform string transitions have been put forward
\cite{Horowitz:2001cz,Gubser:2001ac,Harmark:2002tr,Horowitz:2002dc,Kol:2002xz,Wiseman:2002ti,Harmark:2003fz,Kol:2003ja}
but it is unclear which of these, if any, are correct.%
\footnote{Other recent and related work includes
\cite{Casadio:2000py,Casadio:2001dc,Horowitz:2002ym,DeSmet:2002fv,Kol:2002dr,Sorkin:2002nu,Frolov:2003kd,Emparan:2003sy}.}

The outline of this paper is as follows.
In Section \ref{secnewdiag} we collect the main results of
\cite{Harmark:2003dg}. We start by reviewing the two parameters
of the new phase diagram, the mass $M$ and a new asymptotic quantity
called the relative binding energy $n$. The plot of the known branches
in the $(M,n)$ phase diagram is then given. We also recall
the new Smarr formula of \cite{Harmark:2003dg} and some of its
consequences for the thermodynamics, as well as the Uniqueness
Hypothesis.

In Section \ref{secansatz} we show that any static and neutral
black hole or string on a cylinder can be put into the ansatz
originally proposed in \cite{Harmark:2002tr}, generalizing
the  result of Wiseman \cite{Wiseman:2002ti} for $d=5$.
This is done in two steps by first deriving the three-function conformal
ansatz, and subsequently showing that one can transform to the two-function
ansatz of \cite{Harmark:2002tr}. The latter ansatz was shown
to be completely determined by one function only, after explicitly
solving the equations of motion. We then comment on the boundary
conditions for this one function and their relation to the asymptotic
quantities for black holes/strings on cylinders introduced in
\cite{Harmark:2003dg}.

In Section \ref{1stsec} we use the ansatz of the previous section
to prove that any curve of solutions always obeys
the first law of thermodynamics
\begin{equation}
\delta M = T \delta S + n M \frac{\delta R_T}{R_T} \ .
\end{equation}
Here the last term is due to the pressure in the periodic direction,
and confirms the fact that $n$ is a meaningful physical quantity.
Moreover, for fixed value of the circle radius $R_T$, this
means that $\delta M = T \delta S$ for all solutions in
the $(M,n)$ phase diagram. We analyze the consequences of this
for the black hole/string phase transitions.

In Section \ref{seccopy} we further develop an argument
of Horowitz \cite{Horowitz:2002dc}
that a given non-uniform black string solution can be ``unwrapped'' to give
a new non-uniform black string solution.
We write down explicit transformations for our ansatz
that generate new solutions
and compute the resulting effect on the thermodynamics.
We then explore the consequences of this solution generator
for both the non-uniform black strings as well as the black hole branch.
In particular, we prove the so-called Invertibility Rule.
Under certain assumptions, this rule puts restrictions
on the continuation of Wiseman's branch.

Furthermore, in Section \ref{secscen} we discuss the consequences of the
new phase diagram for the proposals on the phase structure of black holes and
strings on cylinders. We first go through the three existing proposal
for the phase structure, and subsequently introduce three new proposals
for how the phase structure could be.

Finally, in Section \ref{secline} we make the phenomenological
observation that the non-uniform string branch of
Wiseman \cite{Wiseman:2002zc} has a remarkable linear behavior when
plotting the quantity $TS$ versus the mass $M$.

We have the discussion and conclusions in Section \ref{secconcl}.

Two appendices are included providing some more details and
derivations. Appendix \ref{appequi}
contains a refinement of the original proof \cite{Wiseman:2002ti}
that for $d=5$ the conformal
 ansatz can be mapped onto the ansatz proposed in \cite{Harmark:2002tr}.
A useful identity for static perturbations in the ansatz
of \cite{Harmark:2002tr} is derived in Appendix \ref{iddM}.

\section{New type of phase diagram}
\label{secnewdiag}

In this section we review the paper \cite{Harmark:2003dg} and
summarize the formulas and results that are important for this paper.

We begin by reviewing the new phase diagram of \cite{Harmark:2003dg}.
We first define the parameters of the phase diagram.%
\footnote{Similar asymptotic quantities as the ones given
here have  been previously considered
in \cite{Traschen:2001pb,Townsend:2001rg}.}
Consider some localized, static and neutral Newtonian matter on
a cylinder $\R^{d-1} \times S^1$. Let the cylinder space-time have
the flat metric $-dt^2 + dz^2 + dr^2 + r^2 d\Omega_{d-2}^2$. Here
$z$ is the period coordinate with period $2\pi R_T$. We then
consider Newtonian matter with the non-zero components of the
energy momentum tensor being
\begin{equation}
T_{00} = \varrho
\spa
T_{zz} = - b \ .
\end{equation}
We define then the mass $M$ and the {\sl relative binding energy} $n$ as
\begin{equation}
\label{Mndefs}
M = \int d^{d} x \, \rho(x)
\spa
n = \frac{1}{M} \int d^{d} x \, b(x) \ .
\end{equation}
$n$ is called the relative binding energy since it is the
binding energy per unit mass.
We can measure $M$ and $n$ by considering the leading part of
the metric for
\begin{equation}
\label{ctcz}
g_{00} = - 1 + \frac{c_t}{r^{d-3}} \spa
g_{zz} = 1 + \frac{c_z}{r^{d-3}} \spa
\end{equation}
for $r\rightarrow \infty$. In terms of $c_t$ and $c_z$ we have
\begin{equation}
\label{findMn}
M = \frac{\Omega_{d-2} 2\pi R_T}{16 \pi G_{\rm N}} \left[ (d-2) c_t - c_z \right]
\spa
n = \frac{c_t - (d-2) c_z}{(d-2) c_t - c_z} \ .
\end{equation}
Eqs.~\eqref{ctcz}-\eqref{findMn} can then be used to find the mass $M$ and
the relative binding energy $n$ for any black hole or string solution on a
cylinder.

\begin{figure}[ht]
\centerline{\epsfig{file=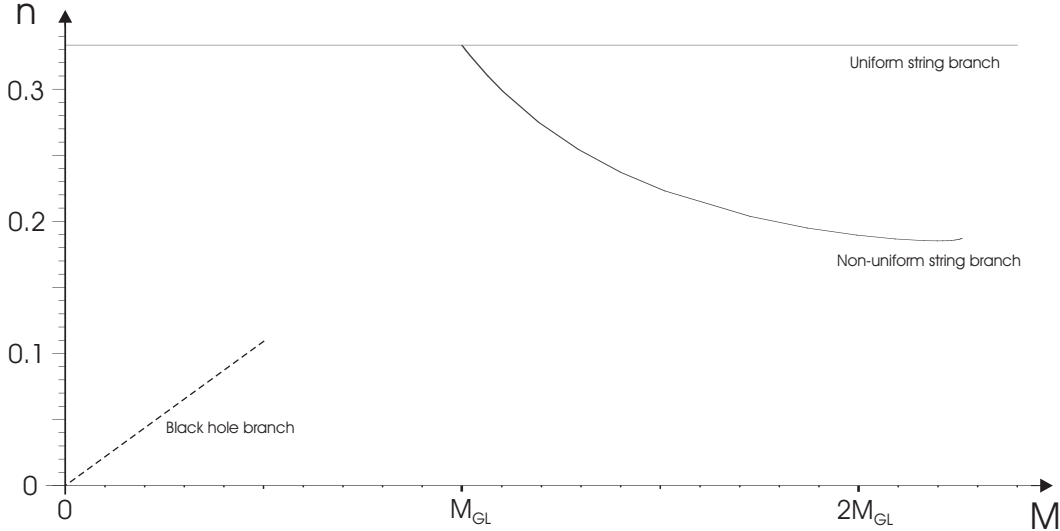,width=14cm,height=7cm}}
\caption{$(M,n)$ phase diagram for $d=5$ containing
the uniform string branch and the non-uniform string branch
of Wiseman. The black hole branch is sketched.}
\label{phase1}
\end{figure}

Three known phases of black holes and strings on cylinders are
depicted in Figure \ref{phase1} for the case $d=5$.
The uniform black string phase has $n=1/(d-2)$ and exists for
all masses $M$.
For the non-uniform black string phase that starts at
the Gregory-Laflamme point $(M,n)=(M_{\rm GL},1/(d-2))$
we have depicted the solutions found by Wiseman in \cite{Wiseman:2002zc}
for $d=5$ (see \cite{Harmark:2003dg} for details on how the data
of \cite{Wiseman:2002zc} are translated into the $(M,n)$ phase diagram).
The black hole branch is sketched in Figure \ref{phase1}.
The branch starts at $(M,n)=(0,0)$ but apart from that the precise
curve is not known.

A useful result for studying the thermodynamics
of solutions in the $(M,n)$ phase diagram is the Smarr formula
\cite{Harmark:2003dg}%
\footnote{In Ref.~\cite{Harmark:2003dg} the numerical data of Wiseman
was used to explicitly check that the Smarr formula holds to very high precision for the
non-uniform black string branch.}
\begin{equation}
\label{Smarra}
TS = \frac{d-2-n}{d-1} M \ .
\end{equation}
A curve of solutions obeying the first law $\delta M = T \delta S$
is called a branch. For branches of solutions, it then follows
from \eqref{Smarra} that
\begin{equation}
\label{intS}
\frac{\delta \log S}{\delta M} = \frac{1}{M} \frac{d-1}{d-2-n(M)} \ ,
\end{equation}
so that given the curve $n=n(M)$, one can integrate
to obtain the entire thermodynamics.

Another important consequence of the Smarr formula is the
Intersection Rule. This states that for two
intersecting branches we have the property
that for masses below an intersection point the branch
with the lower relative binding energy has the highest entropy,
whereas for masses above an intersection point it is the branch
with the higher relative binding energy that has the highest entropy.

In \cite{Harmark:2003dg} we also commented on the uniqueness
of black holes/strings on cylinders.
It seems conceivable that the higher Fourier modes (as defined
in \cite{Harmark:2003dg}) are determined by $M$ and
$n$ for black holes/strings. Therefore we proposed in \cite{Harmark:2003dg}
the following:
\begin{itemize}
\item[$\blacksquare$] {\bf Uniqueness Hypothesis}\newline
Consider solutions of the Einstein equations on a cylinder
$\R^{d-1} \times S^1$.
For a given mass $M$ and relative binding energy $n$
there exists at most one neutral and static solution
of the Einstein equations with an event horizon.
\end{itemize}
It would be very interesting to examine the validity of the
above Uniqueness Hypothesis. If the higher Fourier modes
of the black hole/string instead are independent of $M$ and $n$,
it would suggest
that one should have many more dimensions in a phase diagram
to capture all of the different black hole/string solutions.

\section{Derivation of ansatz for solutions}
\label{secansatz}

\subsection{Derivation of ansatz}
\label{secderiv}

In this section we derive that any neutral and static black
hole/string on a cylinder $\R^{d-1} \times S^1$ (of radius $R_T$)
can be written in a very restrictive ansatz, originally
proposed in \cite{Harmark:2002tr}.
This generalizes an argument of Wiseman in \cite{Wiseman:2002zc}.

Note that it is assumed in the following that any static
and neutral black string or hole on a cylinder $\R^{d-1} \times S^1$
is spherically symmetric in $\R^{d-1}$.

\subsubsection*{The black string case}

We begin by considering the black string case, i.e. where the
topology of the horizon is $S^{d-2} \times S^1$.
The black hole case is treated subsequently.

Using staticity and spherical symmetry in $\R^{d-1}$,
the most general form of the
metric for a black string is
\begin{equation}
\label{fiveans} ds^2 = - e^{2B} dt^2
+ \sum_{a,b=1,2} g_{ab} \, dx^a dx^b + e^{2D} d \Omega_{d-2}^2 \ ,
\end{equation}
where  $B$, $D$, $g_{ab}$ are functions of $x^1$ and $x^2$ only.
This ansatz involves 5 unknown functions, and the
location of the event horizon is given by the relation
$e^{B(x_1,x_2)} = 0$.

Since we are considering black strings wrapped on the cylinder,
i.e. with an event horizon of topology $S^{d-2} \times S^1$, we can
think of the
two coordinates $x^1$ and $x^2$ as coordinates
on a two-dimensional semi-infinite cylinder of topology
$\R_+ \times S^1$ and with metric $g_{ab}$,
where the boundary of the cylinder is the location of the event
horizon of the black string.%
\footnote{We use here and in the following the notation
that $\R_+ = [0,\infty [$.}

Now, it is a mathematical fact that any two-dimensional Riemannian
manifold is conformally flat. We can thus find coordinates $r,z$ so
that
\begin{equation}
\sum_{a,b=1,2} g_{ab} \, dx^a dx^b = e^{2C} (dr^2 + dz^2) \ .
\end{equation}
Since $dr^2 + dz^2$ now describes a Ricci-flat semi-infinite
cylinder of topology $\R_+ \times S^1$ we can choose $z$ as the
periodic coordinate for the cylinder and $r$ to be the coordinate
for $\R_+$, i.e $r=0$ is the boundary of the cylinder and $r \geq
0$.

With this, we have shown that the metric \eqref{fiveans}
can be put in the conformal form
\begin{equation}
\label{confans}
ds^2 = - e^{2B} dt^2 + e^{2C} (dr^2 + dz^2) + e^{2D} d \Omega_{d-2}^2 \ ,
\end{equation}
involving three functions only.
Moreover,
$z$ is a periodic coordinate of period $2\pi R_T$
and $r \geq 0$, with $r=0$ being
the location of the horizon. The form
\eqref{confans} of the black string metric is called the conformal ansatz,
which  was also derived and used
in Refs.~\cite{Gubser:2001ac,Wiseman:2002zc} for $d=4,5$.

We now wish to further reduce the three-function conformal ansatz to
the two-function ansatz of Ref.~\cite{Harmark:2002tr}. For $d=5$
this was done in \cite{Wiseman:2002ti} as reviewed in appendix
\ref{appequi}, where we also add a small refinement.
Here we generalize the argument to arbitrary
$d$. The two-function ansatz  reads
\begin{equation}
\label{2fans}
ds^2 = -f dt^2 + R_T^2 \left[ f^{-1} A d R^2  + \frac{A}{K^{d-2}} dv^2 +
K R^2 d \Omega_{d-2}^2 \right] \spa f = 1 - \frac{R_0^{d-3}}{R^{d-3}} \ ,
\end{equation}
with $A$ and $K$ functions of $R$ and $v$.
We first perform one
coordinate transformation and a redefinition of $A$ and $K$
\begin{equation}
\label{redef}
R^{d-3} = R_0^{d-3} + \bar r^{d-3}
\spa
\hat A = f^{-1}  A R_T^2 \left( \frac{\bar r}{R} \right)^{2(d-4)}
\spa
\hat K^{d-2} = \frac{K^{d-2} }{f} \left( \frac{\bar r}{R} \right)^{2(d-4)}
\ ,
\end{equation}
after which we find
\begin{equation}
ds^2 = - \frac{\bar r^{d-3}}{R_0^{d-3}+\bar r^{d-3}} dt^2 +
\hat A \left[  d \bar r^2  + \frac{1}{\hat K^{d-2}} dv^2 \right] +
R_T^2 \hat K  \hat r^{\frac{5-d}{d-2}} \left( R_0^{d-3} +\hat r^{d-3} \right)^{\frac{3}{d-2}}
d \Omega_2^{d-2} \ .
\end{equation}
We then write $\hat A = e^{2a}$ and $R_T^2\hat K = e^{2k}$ and change to
the conformal form \eqref{confans} by the coordinate transformation
\begin{equation}
\label{gandh}
 \bar r = g(r,z) \spa v = h (r,z) \ ,
\end{equation}
\begin{equation}
\label{coorddnn}
\partial_r g =  e^{-(d-2)k} \partial_z h \spa
\partial_z g = - e^{-(d-2)k} \partial_r h \ .
\end{equation}

Matching the lapse function of the resulting metric with that
in \eqref{confans} gives the condition
\begin{equation}
\label{lapsedn}
g^{d-3} = \frac{R_0^{d-3}  e^{2B}}{1- e^{2B}} \ ,
\end{equation}
while identifying the rest of the metric yields
\begin{equation}
\label{relkdn}
e^{2a} = \frac{e^{2C}}{ (\partial_r g)^2 + (\partial_z g)^2}
\ , \ \
e^{2k} = \frac{R_T^2}{R_0^2}e^{2D} e^{ \frac{2(d-5)}{(d-2)(d-3)}B }
\left( 1-e^{2B} \right)^{\frac{2}{d-3}} \ .
\end{equation}
However, the system \eqref{coorddnn} implies an integrability
 condition on $g(r,z)$ which reads
\begin{equation}
 (\pr^2 + \pz^2) g + (d-2) (\pr g \pr k + \pz g \pz k) = 0 \ .
 \end{equation}
Inserting in this equation the form of $g$ in \eqref{lapsedn}
and $k$ from \eqref{relkdn} results in the second order differential
equation
\begin{equation}
\label{2ordDE}
(\pr^2 + \pz^2 )B + (\pr B)^2 + (\pz B)^2 + (d-2) (\pr B \pr D
+ \pz B \pz D) = 0 \ .
\end{equation}
This turns out to be precisely the
equation of motion $R_{tt} =0$ in the ansatz \eqref{confans}.

As a consequence we have shown that the ansatz \eqref{2fans} is a
{\it consistent} ansatz,  so that in particular, the four
differential equations for the two functions $A(R,v)$ and $K(R,v)$
(see eqs.~(6.1)-(6.4) of \cite{Harmark:2002tr}) are a consistent
set of differential equations,
as was conjectured in \cite{Harmark:2002tr}.%
\footnote{In \cite{Harmark:2002tr} it was shown that the
four equations (6.1)-(6.4) of \cite{Harmark:2002tr} are consistent
to second order in a perturbative expansion of these equations
for large $R$.}

To complete the argument, we also need to consider the location
of the horizon and the periodicity in the compact direction. In the conformal ansatz these are given by $r=0$, and $z \simeq z +2 \pi R_T$ respectively. As far as
the horizon is concerned, it follows immediately from \eqref{lapsedn}
that in the two-function ansatz the horizon will be at
$ \bar r  =0$, as desired (since this means $R=R_0$ by \eqref{redef}).

We moreover want that $A(R,v)$ and $K(R,v)$ are periodic in $v$
with period $2\pi$.%
\footnote{The reason for requiring $2\pi$ as period
is that $A,K \rightarrow 1$
asymptotically and then $v$ has circumference $2\pi R_T$ from
\eqref{2fans}.}
A necessary requirement is that $v \rightarrow v+2\pi$ when
$z \rightarrow z+2\pi R_T$.
{} From \eqref{gandh} we see this is true provided
\begin{equation}
\label{barRT}
h(r, z + 2 \pi R_T ) - h(r,z) = 2\pi \ .
\end{equation}
We first prove that the left hand side of \eqref{barRT} is independent of $r$.
This is proven by noting that since $g(r,z)$ is periodic in $z$
by \eqref{lapsedn} then $\partial_z g$ is periodic in $z$ and
from \eqref{coorddnn} it follows that then $\partial_r h$ is
also periodic in $z$, which proves the statement.

We then write the left hand side of \eqref{barRT} as
\begin{equation}
\label{delzh}
\int_{0}^{2 \pi R_T} dz \, \partial_z h =
\int_{0}^{2 \pi R_T} dz \, e^{(d-2)k} \partial_r g \ ,
\end{equation}
where we used \eqref{coorddnn} in the last step.
Since we know that \eqref{delzh} is independent of $r$ we may
evaluate the integrand for $r \rightarrow \infty$.
We know that $k \rightarrow 0$ in this limit, so this
leaves $\partial_r g$.
However, from \eqref{ctcz} we see that $1-e^{2B} = c_t r^{-(d-3)}$
to leading order so using
\eqref{lapsedn} we finally get from \eqref{delzh} that \eqref{barRT}
is true provided that $c_t = R_0^{d-3}$, which also follows from
\eqref{2fans}
since $R/r \rightarrow 1$ asymptotically.

We can think of the coordinate transformation from
$(r,z)$ to $(R,v)$ as a map from $\R_+ \times S^1$
to $\R_+ \times S^1$. This map is non-singular and invertible
since we are dealing with the black string case where
the curves $\{ R = \mbox{constant} \}$ have $S^1$ topology in the
$\R_+ \times S^1$ cylinder given by the $(r,z)$ coordinates.
Thus, that $\{ R = \mbox{constant} \}$ has $S^1$ topology means
that it has a periodicity in $v$, and in order for the
total coordinate map to be continuous, the periodicity
must be the same everywhere. Hence, since the periodicity is
$2\pi$ asymptotically, it is $2\pi$ everywhere.
We therefore conclude that $A(R,v)$ and $K(R,v)$ are
periodic functions in $v$ with period $2\pi$.

\subsubsection*{The black hole case}

The proof that any static and neutral
black hole on a cylinder $\R^{d-1} \times S^1$
can be written in the ansatz \eqref{2fans} follows basically
along the same lines as the black string argument above,
though with some notable differences.

Any static and neutral black hole on a cylinder $\R^{d-1} \times S^1$ can be
written in the ansatz \eqref{fiveans}. The location of the event horizon
is still specified by $e^{B} = 0$. Topologically we can again think of
$(x^1,x^2)$ as being coordinates on a semi-infinite cylinder
$\R_+ \times S^1$ with metric $g_{ab}$.
The metric $g_{ab}$, $a,b=1,2$, for the two-dimensional cylinder is
non-singular away from the boundary $e^B = 0$ and we can get
\eqref{confans} again by using that any two-dimensional Riemannian
manifold is conformally flat. We furthermore  require again that $z$ is
periodic with period $2\pi R_T$.
Using the same coordinate transformations \eqref{redef}-\eqref{coorddnn} as
for the string case
we find that \eqref{confans} can be transformed into \eqref{2fans}
since the equation of motion $R_{tt} = 0$ again proves the consistency of the
transformation.
{}From \eqref{lapsedn} it is furthermore evident that the horizon
for the black hole is at $R=R_0$, as it should.

In conclusion, we have proven that any static and neutral black
hole solution on a cylinder $\R^{d-1} \times S^1$
can be written in the ansatz \eqref{2fans}.

The only remaining issue is the periodicity in $v$.
Clearly, we still have that $v \rightarrow v+2\pi$ when
$z \rightarrow z+2\pi R_T$. But, the coordinate map from $(r,z)$
to $(R,v)$ is not without singularities.
The qualitative understanding of this is basically that
the curves $\{ R = \mbox{constant} \}$ still have topology $S^1$
in the $(r,z)$ coordinates when $R$ is sufficiently large.
But for small $R$ the curve $\{ R = \mbox{constant} \}$
ends on the boundary of $\R_+ \times S^1$. However,
as explained in \cite{Harmark:2002tr} for a specific
choice of $R$ and $v$ coordinates, the functions
$A(R,v)$ and $K(R,v)$ are still periodic in $v$ with
period $2\pi$ for small $R$.

\subsection{Ansatz for black holes and strings on cylinders}

Assuming spherical symmetry on the $\R^{d-1}$ part
of the cylinder, we have derived in Section \ref{secderiv} that the
metric of any static and neutral black hole/string on a cylinder
can be put in the form
\begin{equation}
\label{ansatz}
ds^2 = - f dt^2 + R_T^2 \left[ f^{-1} A dR^2 +
\frac{A}{K^{d-2}} dv^2 + K R^2 d\Omega_{d-2}^2 \right] \ , \ \
f = 1 - \frac{R_0^{d-3}}{R^{d-3}} \ ,
\end{equation}
with $A = A(R,v)$ and $K=K(R,v)$.
This ansatz was proposed and studied in \cite{Harmark:2002tr}
as an ansatz for black holes on cylinders.

The ansatz \eqref{ansatz} has two functions $A(R,v)$
and $K(R,v)$ to determine. However, in \cite{Harmark:2002tr}
it was shown that $A(R,v)$ can be written explicitly in terms
of $K(R,v)$ and its partial derivatives
(see eq.~(6.6) of \cite{Harmark:2002tr}). Thus, the ansatz
\eqref{ansatz} is completely determined by only one function.

We impose the following boundary condition on $K(R,v)$,
\begin{equation}
\label{Kcond}
K(R,v) = 1 - \frac{1-(d-2)n}{(d-3)(d-2-n)} \frac{R_0^{d-3}}{R^{d-3}}
+ \cdots  \ .
\end{equation}
which implies, using the equations of motion \cite{Harmark:2002tr},
the same boundary condition on $A(R,v)$.
Note that the fact that $K \rightarrow 1$ for $R_0/R \rightarrow 0$
means that $R = r/R_T$ and $v = z/R_T$ when $R \rightarrow \infty$.
Moreover, the condition \eqref{Kcond} ensures that
we have the right asymptotics at $r \rightarrow \infty$, according
to Section \ref{secnewdiag}. We can then use \eqref{Kcond} in
\eqref{findMn}, to compute the mass in terms of $R_0$ and $n$
\begin{equation}
\label{massa}
M = \frac{\Omega_{d-2} 2\pi R_T}{16 \pi G_{\rm N}} ( R_0 R_T )^{d-3}
\frac{(d-1)(d-3)}{d-2-n} \ ,
\end{equation}
which means we can find $R_0$ as function of $M$ and $n$.

Define now the extremal function
\begin{equation}
K_0 (R,v) \equiv K(R,v)|_{R_0 = 0} \ .
\end{equation}
Since $r^2 = K R^2$ and $g_{00} = -1 + R_0^{d-3}/R^{d-3}$,
we then notice that we can write
\begin{equation}
\label{findK0}
\frac{R_0^{d-3}}{r^{d-3}} K_0(r,z)^{\frac{d-3}{2}} = -2 \, \Phi (r,z)
- \frac{2}{(d-1)(d-3)} \frac{8\pi G_{\rm N}}{\Omega_{d-2} 2\pi R_T}
\frac{nM}{r^{d-3}} \ ,
\end{equation}
with $\Phi(r,z)$ given by
\begin{equation}
\Phi (r,z) = - \frac{d-2}{(d-1)(d-3)}\frac{8\pi G_{\rm N}}{2\pi R_T}
\sum_{k=0}^{\infty} \frac{1}{r^{d-3}}
h \! \left( \frac{kr}{R_T} \right) \cos \left( \frac{kz}{R_T}
\right) \varrho_k \ .
\end{equation}
Here $\varrho_k$ are the Fourier modes defined in \cite{Harmark:2003dg},
which characterize the $z$ dependence of the black hole/string.
Therefore, we see that given $M$, $n$ and $\varrho_k$, $k\geq 1$,
we obtain $K_0(r,z)$ and thereby $K_0(R,v)$.
Thus, imposing the two boundary conditions \eqref{Kcond} and \eqref{findK0}
corresponds to imposing specific values for $M$, $n$ and $\varrho_k$.

As shown in \cite{Harmark:2002tr}, one of the nice features of the ansatz
\eqref{ansatz} is that it has a Killing horizon at $R=R_0$
with a constant temperature, provided $K(R,v)$ is finite for
$R=R_0$. Thus, unless $K(R,v)$ diverges for $R \rightarrow R_0$, the
ansatz corresponds to a black hole or string.
For use below, we recall here the corresponding expressions for
temperature and entropy \cite{Harmark:2002tr}
\begin{equation}
\label{neutTS}
T =  \frac{d-3}{4\pi \sqrt{A_h} R_0 R_T}
\spa
S =  \frac{\Omega_{d-2} 2\pi R_T \sqrt{A_h}}{4G_{\rm N}}
(R_0 R_T)^{d-2} \ ,
\end{equation}
where $A_h \equiv A(R_0,v)$.

\section{First law of thermodynamics \label{1stsec}}

In this section we derive the first law of thermodynamics
for static and neutral black holes/strings on a cylinder, and
discuss its consequences for the $(M,n)$ phase diagram.

\subsection{Proof of the first law}

The three central ingredients of our derivation of
the first law of thermodynamics are i) The Smarr law
\eqref{Smarra}
 ii) the identity derived in appendix \ref{iddM}
and iii) the ansatz \eqref{ansatz} describing black holes/strings
on the cylinder.

We first note that by variation of
the Smarr law \eqref{Smarra} we find the relation
\begin{equation}
\label{difSmarr}
S \de T  + T \de S = \frac{d-2-n}{d-1} \de M - \frac{1}{d-1} M \de n \ .
\end{equation}

Our aim is now to use the identity \eqref{walds}, \eqref{vwdef}
to obtain another differential relation involving $\de M$.
To this end we first rescale the ansatz \eqref{ansatz} so that
it becomes
\begin{equation}
\label{reans}
ds^2 = - f dt^2 + R_T^2 \left[ R_0^2 f^{-1} A dR^2
+ \frac{A}{K^{d-2}} dv^2 + R_0^2 K R^2 d \Omega_{d-2}^2 \right]
\ , \
f = 1 - \frac{1}{R^{d-3}} \ .
\end{equation}
In this way of writing the ansatz the horizon is always located at
$R=1$.

Consider now a given time $t=t'$ and take $S_1$
to be the $(d-1)$-dimensional null surface $S_h$ defined $t=t'$ by
$R=1$ and $S_2$ as the $(d-1)$-dimensional space-like
surface $S_{\infty}$ specified by $t=t'$ and $R = R_m$ with
$R_m \rightarrow \infty$.
Putting this into the identity \eqref{walds} along with the metric
\eqref{reans} we get
\begin{equation}
\label{waldt}
\int_{v=0}^{2\pi} dv \, A(R=1,v) \, \Gamma (R=1,v)
= R_m^{d-2} \int_{v=0}^{2\pi} dv \, A(R=R_m,v) \, \Gamma (R=R_m,v) \ ,
\end{equation}
\begin{equation}
\Gamma(R,v) = (g^{RR})^2 \partial_R \delta g_{RR}
- g^{RR} g^{\mu \nu} \Gamma^\alpha_{\mu R} \delta g_{\alpha \nu}
- g^{RR} g^{\mu \nu} \Gamma^{R}_{\mu \nu} \delta g_{RR}
- g^{RR} \partial_R ( g^{\mu \nu} \delta g_{\mu \nu} ) \ .
\end{equation}
We now use the rescaled ansatz \eqref{reans} to compute $\Gamma(R,v)$
and substitute
this into \eqref{waldt}. After some algebra we obtain the
relation
\begin{equation}
\label{Wald2}
 \frac{\delta R_0}{R_0}
+ \frac{1}{2} \frac{\delta A_h}{A_h}
= \frac{(d-3)(1+n)}{d-2-n} \frac{\delta R_0}{R_0}
+ \frac{d-1}{(d-2-n)^2} \delta n \ ,
\end{equation}
where $A_h= A(R=1,v)$. We recall here that it follows from
 the equations of motion that this quantity is constant
on the horizon \cite{Harmark:2002tr}.
To obtain the left hand side we have expressed the
variation of
the metric on the horizon in terms of $\delta R_0$ and
$\delta A_h$ (the variation $\delta K$ does not appear, nor
does the quantity $K(R=1,v)$ which is
generally not constant on the horizon).
The right hand side is obtained by expressing the variation of the
metric at infinity in terms of $\delta R_0$ and $\delta n$, where we have
used the asymptotic form \eqref{Kcond} for $K(R,v)$ and the same
form of $A(R,v)$ as is required by the equations of motion.

To reexpress the identity \eqref{Wald2} in terms of thermodynamic
quantities we use the expressions \eqref{neutTS} for the temperature $T$ and entropy $S$ of the ansatz \eqref{ansatz} along with the
mass in \eqref{massa}.
The identity \eqref{Wald2} can then be put in the form
\begin{equation}
-S \, \de T = \frac{1+n}{d-1} \, \de M + \frac{1}{d-1}   M \delta n \ .
\end{equation}
Adding this to the differential relation \eqref{difSmarr} that followed
from Smarr, we immediately obtain
\begin{equation}
\de M = T \de S \ ,
\end{equation}
which is the first law of thermodynamics.%
\footnote{Note that indeed the relation in eq.(7.13) of
\cite{Harmark:2002tr}, which was shown to hold iff the first law
holds, coincides precisely with the relation \eqref{Wald2} when
using $\gamma = 1/\sqrt{A_h}$ and $\chi =
\frac{1-(d-2)n}{(d-3)(d-2-n)}$.}
The first law indeed holds for all known branches in the phase
diagram. For the uniform black string branch and small black holes
on the cylinder this is of course obvious. For the non-uniform
black string branch, one can also verify the validity to
high precision (see Appendix C of \cite{Harmark:2003dg}).

So far we have kept the radius $R_T$ of the periodic direction
fixed. It is a simple matter to repeat the analysis above by
also allowing $R_T$ to vary. The result is that the first law
acquires an extra work term and becomes
\begin{equation}
\label{gen1st}
\de M = T \de S + nM \frac{\delta R_T}{R_T} \ .
\end{equation}
The extra term could in fact be expected on general physical
grounds. This is because from \eqref{Mndefs} we see
that the average pressure is $p = - n M / V$ and since
$\delta R_T / R_T = \delta V / V$ we have
$nM \delta R_T / R_T = - p \delta V$, giving
$\de  M = T \de S - p \delta V$.
This observation also confirms the fact that $n$ really is a meaningful
physical quantity and it provides a nice
thermodynamic consistency check on the connection
between $n$ and the negative pressure $p$ in the periodic direction.

\subsection{Consequences for space of solutions}

We have proven above that {\sl any} curve
of solutions in the $(M,n)$ diagram has to obey the
first law $\delta M = T \delta S$.
The most important consequence of this is that there
cannot exist solutions for all masses $M \geq 0$ and
relative binding energies $0 \leq n \leq 1/(d-2)$.
This is perhaps surprising, since using the ansatz \eqref{ansatz}
one can easily come up with a solvable system of equations
and asymptotic boundary conditions%
\footnote{One can for example use the asymptotic boundary conditions
described in \cite{Harmark:2002tr}.}
(at $R \rightarrow \infty$)
that would correspond to a given $M$ and $n$.
The resolution of this is of course then that the horizon is only
regular for special choices of $M$ and $n$. In terms of the ansatz
\eqref{ansatz} this means that only for special choices of $M$ and $n$
we have that $A(R,v)$ and $K(R,v)$ are finite on the horizon $R=R_0$.%
\footnote{Note that this shows that the mechanism proposed in
\cite{Harmark:2002tr} to choose which curve in the $(M,n)$ diagram
corresponds to the black hole branch is then proven wrong, since
in that paper it was proposed that one could use the
first law $\delta M = T \delta S$ to find the black hole branch.}

The above proof that any curve of solutions obeys the first law
moreover makes it highly doubtful that there is any set of
solutions with a two-dimensional measure in the $(M,n)$ phase diagram.
Thus, for any two given branches of solutions, we do not expect there
to be a continuous set of branches in between them. The main reason for this
is that if one had two different points in the
$(M,n)$ phase diagram with same mass, the first law $\delta M = T \delta S$
would imply they had the same entropy
since it would be possible to go from one point to the other varying only $n$.
In fact, it is clearly true that
the uniform branch, the non-uniform branch that is
connected to the Gregory-Laflamme point $(M,n)=(M_{\rm GL},1/(d-2))$
and the black hole branch, cannot be continuously connected in
this way.%
\footnote{The same is true for  any two copies (as defined in
Section \ref{seccopy}) of the non-uniform branch connected to the
Gregory-Laflamme point $(M,n)=(M_{\rm GL},1/(d-2))$.}

For the Horowitz-Maeda conjecture  \cite{Horowitz:2001cz}
that there exist light non-uniform strings with entropy
larger than that of a uniform string of equal mass,
this means that those solutions should be a one-dimensional
set of solutions which describe the end point of the
decay of a light uniform string. The intermediate classical
solutions that the classical decay of the uniform string evolves through
(see \cite{Choptuik:2003qd} for numerical studies of the decay)
should therefore be non-static.

\section{Copies of solutions}
\label{seccopy}

\subsection{Solution generator}

In \cite{Horowitz:2002dc} Horowitz argued that for any non-uniform
string solution one can ``unwrap'' this solution to get a new solution
which is physically different. The argument is essentially that
if one takes a solution with mass $M$ and circumference $L$ of a
cylinder $\R^{d-1} \times S^1$, then one can change the
periodicity of the cylinder coordinate $z$ from $L$ to
$kL$, with $k$  a positive integer. This will produce a new solution
with mass $\tilde{M} = kM$ and cylinder circumference $\tilde{L} = kL$.
Since $M/L^{d-2}$ is
dimensionless, we see that the new solution has
$\tilde{M}/\tilde{L}^{d-2} = k^{-(d-3)} M/L^{d-2}$.
In the following we first write down an explicit ``solution generator''
for our ansatz \eqref{ansatz} that directly implements Horowitz's
``unwrapping'' and we then explore the consequences for
black strings and black holes in  Sections
\ref{secstrcopy} and \ref{secn0} respectively.

The specific transformation that
takes any black hole/string solution on the cylinder and
transforms it to a new black hole/string solution on the cylinder
with the same radius, is as follows:

\begin{itemize}
\item[$\blacksquare$]
Consider a black hole/string solution on a cylinder with radius
$R_T$ and horizon radius $R_0$,
and with given functions $A(R,v)$ and $K(R,v)$ in \eqref{ansatz}.
Then for any $k \in \{ 2,3,... \}$ we obtain a new solution as
\begin{equation}
\label{newsol}
\tilde{A}(R,v) = A(kR,kv) \ , \ \
\tilde{K}(R,v) = K(kR,kv) \ , \ \
\tilde{R}_0 = \frac{R_0}{k} \ .
\end{equation}
This gives a new black hole/string solution on a cylinder of
the {\sl same} radius.
\end{itemize}

The statement above may be verified explicitly from the
equations of motion (see eqs.~(6.1)-(6.4) of \cite{Harmark:2002tr}).
One can also easily implement this transformation in the ans\"atze
\eqref{fiveans} and \eqref{confans} instead, but we choose here
for simplicity
to write the transformation in the ansatz \eqref{ansatz}.%
\footnote{In the ansatz \eqref{confans} the transformation
is instead $\tilde{B}(r,z) = B(kr,kz)$,
$\tilde{C}(r,z) = C(kr,kz)$ and $\tilde{D}(r,z) = D(kr,kz)$.}

We can now read off from \eqref{ctcz}, \eqref{findMn}, \eqref{massa}
and \eqref{neutTS} that the new solution \eqref{newsol}
has mass, relative binding energy, temperature and entropy given
by
\begin{equation}
\label{transf}
\tilde{M} = \frac{M}{k^{d-3}} \spa
\tilde{n} = n \spa
\tilde{T} = k T \spa
\tilde{S} = \frac{S}{k^{d-2}} \ .
\end{equation}
We see that the transformation rule for the mass is precisely that
of Horowitz in \cite{Horowitz:2002dc}. Note also that the rule is
in agreement with  the Smarr formula \eqref{Smarra}.

\subsection{Consequences for non-uniform black strings}
\label{secstrcopy}

We can now use the transformation \eqref{transf} on the
known branches of solutions.
For the uniform string branch we merely go from one point
in the branch to another, i.e. the transformation does not
create any new branch of solutions. In fact, using the
entropy expression for the uniform black string
\begin{equation}
S_{\rm BS} (M) = d_1 M^{\frac{d-2}{d-3}} \spa
d_1 = 4 \pi (d-3)^{-\frac{d-2}{d-3}}
\left( \frac{16 \pi G_{\rm N}}{\Omega_{d-2} 2 \pi R_T} \right)^{\frac{1}{d-3}}
\ ,
\end{equation}
it is easy to verify from \eqref{transf} that the entropy of
the $k$'th copy is the same as that of the original branch.

\begin{figure}[ht]
\centerline{\epsfig{file=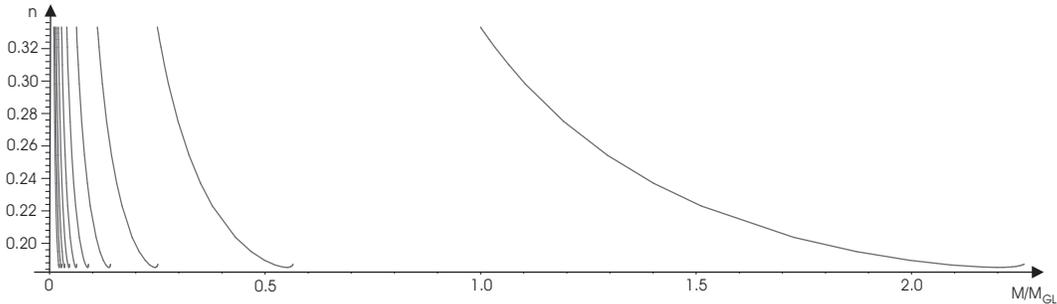,width=14cm,height=4cm}}
\caption{Part of the $(M,n)$ phase diagram for $d=5$
with copies of the non-uniform string branch of Wiseman.}
\label{copies}
\end{figure}

Considering instead the non-uniform string branch that is connected
to the Gregory-Laflamme point $(M,n)=(M_{\rm GL},1/(d-2))$
we see that we do create new branches, in fact infinitely many
new branches. We have depicted this in Figure \ref{copies}
using Wiseman's data for $d=5$.

Note that by the Intersection Rule \cite{Harmark:2003dg} reviewed
in Section \ref{secnewdiag}
all the copies  in Figure \ref{copies} also have the
property that the entropy for a solution is less than that
of a uniform string of same mass.
If we take two copies which both exist for the same
mass, we furthermore see from the Intersection Rule that
the branch which lies to the left in the $(M,n)$ diagram will
have less entropy for a given mass.

We are now capable of addressing an important question regarding
Wiseman's branch (see Figure 5 of \cite{Harmark:2003dg}
for a more detailed plot):
Is Wiseman's branch really starting to have increasing
$n$ at a certain point?
This question can be addressed using the following rule that we prove below.

\begin{itemize}
\item[$\blacktriangleleft$] {\bf Invertibility Rule}\newline
Consider a curve of solutions in the $(M,n)$ phase diagram.
There are no intersections between the curve and its copies,
or between any of the copies, if and only if the curve
is described by a well-defined function $M=M(n)$ (i.e.
if a given $n$ only corresponds to one $M$).
\end{itemize}

We call this the Invertibility Rule since we normally would
describe a branch by a function $n=n(M)$, and then the Invertibility
Rule implies such a function is invertible.

For the non-uniform string branch in Figure \ref{phase1} the Invertibility Rule
implies that the small increase in $n$ in the end of the branch
can be ruled out if we also assume the {\sl Uniqueness Hypothesis} \cite{Harmark:2003dg}
(recalled in Section \ref{secnewdiag}) to be valid.
To see this note that the latter assumption implies that an intersection
of two branches means in particular that the two branches intersect in the
same solution and not just in the same point in the $(M,n)$ diagram.
Thus, using the Intersection Rule (see Section \ref{secnewdiag})
we cannot have two branches intersecting twice, and
if the non-uniform branch was not invertible, then they clearly
would intersect twice since they already both intersect
the uniform string branch. It then follows that the non-uniform branch must
continue to decrease in $n$, and the small increase that one can see must
be from inaccuracies in the numerically computed data.

Although it seems reasonable to assume that the above stated conditions hold,
it would be interesting to further examine the behavior of the Wiseman
branch at the end.

We conclude by proving the Invertibility Rule.
On the one hand,
it is clearly true that a curve described by a well-defined function
$M=M(n)$ will not have intersections between the curve and its copies,
or between any of the copies.
On the other hand,
consider a continuous curve in the $(M,n)$ phase diagram that
is not invertible, so that there are two points $(M_1,n')$ and $(M_2,n')$,
$M_1 < M_2$.
Consider now two copies of the curve according to the
transformation \eqref{transf}, one with $k=p$ and the other with $k=q$.
Then those two copies intersect if
$M_2/p^{d-3} > M_1/q^{d-3}$.
Thus, we have intersections if we can find integers $p$ and $q$
so that
$M_2/M_1 > p^{d-3}/q^{d-3} > 1$. But that is always possible.
This completes the proof.

\subsection{Consequences for black hole branch}
\label{secn0}

We finally consider the consequence of the transformation \eqref{transf} for
black holes on the cylinder, i.e. the branch that starts at $(M,n)=(0,0)$.

We first comment on what, according to \eqref{newsol} and \eqref{transf},
 the physical meaning is of the $k$'th copy for the black hole branch.
{} From \eqref{newsol} it is easy to see that the $k$'th copy of a black hole
is $k$ black holes on a cylinder put at equal distance from each other.
While these solutions clearly exist they are not classically
stable. If we take $k=2$, for instance, it is clear that this solution
consisting of two black holes is unstable against perturbation of the relative
distance between the black holes. The solution will then obviously decay to a
single black hole on a cylinder. More generally, $k$ black holes on a
cylinder will decay to a fewer number of black holes when perturbed.
Thus, all these solutions with $k>1$ are in an unstable equilibrium.
At the level of the entropy this implies that we expect that
$S_{{\rm BH},k} (M) < S_{{\rm BH},k'} (M)$ for $k > k '$.

To see this explicitly, note that the
 thermodynamics for the small black hole branch is given by
\begin{equation}
\label{SBH}
S_{\rm BH} (M) = c_1 M^{\frac{d-1}{d-2}} \spa
c_1 = 4 \pi (d-1)^{-\frac{d-1}{d-2}} \left( \frac{16 \pi G_{\rm N}}{\Omega_{d-1}} \right)^{\frac{1}{d-2}} \ ,
\end{equation}
since it is equal to that of a black hole in $d+1$ dimensional Minkowski space. We can then
use \eqref{transf} and \eqref{SBH} to compute that
the entropy function of the $k$'th copy of the small black hole branch is
\begin{equation}
S_{\rm BH,k} (M) = c_k M^{\frac{d-1}{d-2}} \spa c_k = c_1 k^{-\frac{1}{d-2}}
\ .
\end{equation}
Since then $c_k > c_{k'}$ for $k < k'$, we get that
$S_k(M) > S_{k'}(M)$ for $k < k'$, as we required.

Note that it is not excluded that $n=0$ for all of the black
hole branch. This would even fit well intuitively with the fact that
\eqref{gen1st} in that case tells us that the black hole behaves
like a point-like object since $\delta M=0$ when varying $R_T$.
On the other hand it also seem reasonable to expect that the
complicated shape of the event horizon of the black hole should have
the consequence that there should be some corrections to the flat
space thermodynamics.
Thus, determining whether $n=0$ or
$n>0$ amounts to determining which intuition is right: That the behavior
of black holes as objects are given by the shape of the singularity
or the shape of the event horizon.

Note also that if $n=0$ for the entire black hole branch
the {\sl Uniqueness Hypothesis}
\cite{Harmark:2003dg} which is recalled in Section \ref{secnewdiag}
must be wrong since then we have an infinite number of different
solutions with the same $M$ and $n$.

\section{Possible scenarios and phase diagrams}
\label{secscen}

\subsection{Review of three proposed scenarios}

As mentioned in the introduction,
Horowitz and Maeda argued in \cite{Horowitz:2001cz}
that a classically unstable
uniform black string cannot become a black hole in a classical evolution.
More precisely they showed that an event horizon cannot have a
circle that shrinks to zero size in a finite affine parameter. This led
them to conjecture that there exists a new branch of solutions
which are non-uniform neutral black strings with entropy greater
than that of the uniform black string with equal mass.

The conjecture of Horowitz and Maeda motivated Gubser in
\cite{Gubser:2001ac} to look for a branch of non-uniform black
string solutions, connected to the uniform black string solution
in the Gregory-Laflamme critical point where $M = M_{GL}$, by
using numerical techniques.
The numerical data of Gubser were later improved by
Wiseman in \cite{Wiseman:2002zc}.%
\footnote{Note that in \cite{Gubser:2001ac} the case $d=4$ was
studied while in \cite{Wiseman:2002zc} it was the case $d=5$ instead.}
As mentioned above, we depicted the non-uniform branch
in the $(M,n)$ phase diagram of Figure \ref{phase1}, in the case $d=5$.

However, as noted by \cite{Gubser:2001ac}, the piece of the
non-uniform branch found by \cite{Gubser:2001ac,Wiseman:2002zc}
cannot be the conjectured Horowitz-Maeda non-uniform solutions.
Firstly, it was shown in both \cite{Gubser:2001ac} and \cite{Wiseman:2002zc}
that the entropy of a non-uniform black string on the branch
is smaller than that of a uniform
black string of the same mass.
Secondly, all the new solutions in the branch have
masses bigger than the Gregory-Laflamme mass $M_{GL}$. Clearly
we need non-uniform solutions with masses $M < M_{GL}$ in order
to get the conjectured Horowitz-Maeda non-uniform black strings.

In \cite{Gubser:2001ac} an ingenious proposal was put forward
to remedy this situation.
The idea was that the branch should continue and at some point have
decreasing mass so that we get solutions with $M < M_{GL}$.
Moreover, the entropy on the branch should start becoming higher
than that of the corresponding uniform black string with equal mass.
If true, it means that the classically unstable uniform string
decays classically to the classically stable non-uniform
string with the same mass via a first order transition.
After this, the non-uniform string decays quantum mechanically
to a black hole via another first order transition.

Gubser's scenario obviously does not address the issue of what happens when one increases the mass of a black hole on a cylinder.
One possibility is that the black hole branch simply terminates
when the mass reaches the critical value when the black horizon
reaches all around the cylinder.
Clearly then it should be so that the entropy of a black hole just
below the critical mass should be less than that of the uniform black
string, so that one can have a first order transition from the black
hole to a uniform string of the same mass.
In the following we refer to this combined scenario as ``Scenario I''.

In \cite{Harmark:2002tr} it was instead suggested that the black hole phase
should go directly into a non-uniform string phase which connects
to the uniform string phase at $M = \infty$, i.e.
$n \rightarrow 1/(d-2) \mbox{ for } M \rightarrow \infty$.
Alternatively the branch could instead approach some other limiting value,
i.e. $n \rightarrow n_{\star} \mbox{ for } M \rightarrow \infty$.
In this type of combined scenario we thus have two non-uniform black string
branches, one that connects to the black hole branch and another that
connects to the Gregory-Laflamme point $(M,n)=(M_{\rm GL},1/(d-2))$.
We refer to this type of scenario as ``Scenario II''.

An entirely different scenario was proposed
by Kol in \cite{Kol:2002xz}, where it was conjectured that
the non-uniform string branch found in \cite{Gubser:2001ac,Wiseman:2002zc}
should be continued all the way to the black hole branch.
Since this scenario requires for small masses
that the non-uniform black string branch changes topology and becomes the black hole branch, it would in any case not be able to account for the
conjectured Horowitz-Maeda non-uniform black strings for small
masses.
Therefore, the unstable uniform string should instead decay classically
in infinite ``time'', i.e. affine parameter on the horizon, and
become arbitrarily near the black hole branch but never reach it
classically.
We call Kol's scenario ``Scenario III'' below.

A final relevant issue is whether the non-uniform  branch is classically
stable or unstable, which is presently not known. However, since these
solutions were found by applying numerical methods that work by perturbing a solution which then relaxes to a new fixed point, it is likely that the Wiseman branch is classically stable.%
\footnote{We thank Roberto Emparan for this observation.}
In this connection it is also worth considering the copies of the
Wiseman branch discussed in Section \ref{secstrcopy}.
On the one hand, the fact that for increasing number $k$ the entropy
at a given mass decreases, could suggest that the copies are classically
unstable. On the other hand,  
any pair of copied branches has a different mass range,
as can be seen in Figure \ref{copies}. Moreover, contrary to the
black hole copies which constitute separate objects, in this case
all copies of non-uniform strings have the same topology.
As a consequence we view the classical stability of the Wiseman copies as
another open problem.

\subsection{Phase diagrams for the three scenarios}

We can now try to draw how scenarios I-III should look in the
$(M,n)$ phase diagram for $d=5$.

When drawing the scenarios in the following we use the $(M,n)$
phase diagram in Figure \ref{phase1} as the starting point.
However, in that diagram (see also appendix C of
\cite{Harmark:2003dg}) we see that the non-uniform branch starts
having increasing $n$, just before it ends. We argued in Section
\ref{seccopy} that, under reasonable assumptions,
the non-uniform branch in fact has to continue
decreasing in $n$. The small increase one can see in Figure
\ref{phase1} could thus possibly be due to inaccuracies of the numerics. In
the following we have therefore ignored these last few data points and
assumed that $n$ keeps decreasing.

It is also important to note that when we construct the $(M,n)$ phase
diagrams below we are following the Intersection Rule
\cite{Harmark:2003dg} reviewed in Section \ref{secnewdiag} which,
among other things, has the consequence that one cannot have two
branches intersecting twice, at least not when the branches are
described by well-defined functions $n=n(M)$.

\begin{figure}[ht]
\centerline{\epsfig{file=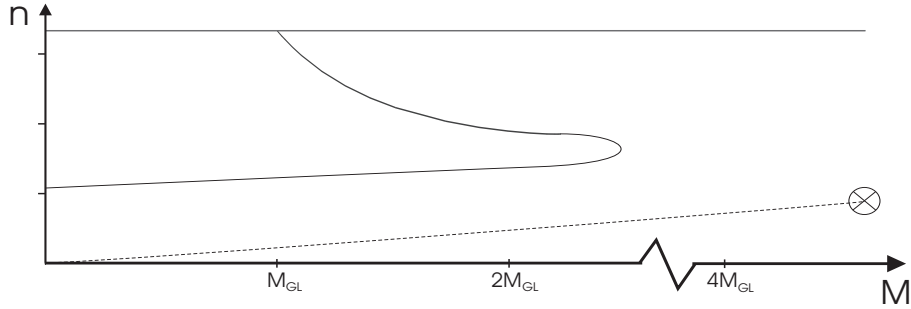,width=12cm,height=4cm}}
\caption{\small Scenario I. Gubser's proposal for the non-uniform
string branch plus a black hole branch that terminates at the
critical point.}
\label{scenI}
\end{figure}

In Scenario I that we described above
we clearly need to let the non-uniform string branch
turn around and reach $M=0$ with $n > 0$ since we do not want to cross
the black hole branch, and we need $n \geq 0$.
We have sketched this scenario in Figure \ref{scenI}.
It looks though unlikely that the curve should make such a complicated
turn, perhaps one can imagine that the curve has a cusp.
This would, however, imply at least a second order
phase transition in the point of the cusp which would seem peculiar
since the solution does not change phase.

In Figure \ref{scenI} we have also depicted
the black hole phase which in Scenario I
terminates at a critical mass when the black hole horizon reaches
around the horizon. Note that since we require the entropy at the
critical point to be higher than that of a uniform string of
the same mass, and since
one can integrate up the entropy \cite{Harmark:2003dg} given a curve in the
$(M,n)$ diagram (see eq.~\eqref{intS}), we can in fact put bounds on the curve.%
\footnote{In particular, this enables us to obtain a lower bound
on the critical mass by equating the entropy of a black hole in
flat Minkowski space to that of a black string. Using the GL masses
listed in Table 1 of \cite{Harmark:2003dg} this yields for $ 4 \leq d \leq 9$
the values $M_{\rm c}/M_{\rm GL} = $
$4.23,\ 4.53, \ 4.09, \  3.98,\ 3.93,\ 3.66$.}

\begin{figure}[ht]
\centerline{\epsfig{file=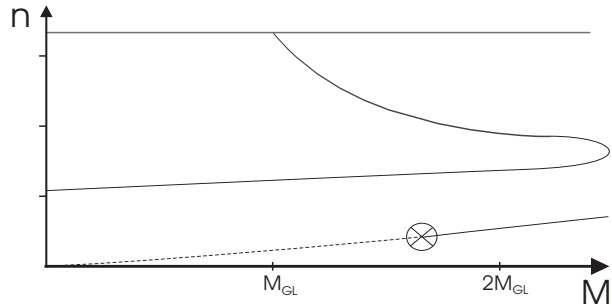,width=8cm,height=4cm}}
\caption{\small Scenario II. Gubser's proposal for the non-uniform
string branch plus a black hole branch that continues into
another non-uniform string branch which continues to arbitrarily
large masses.}
\label{scenII}
\end{figure}
In Figure \ref{scenII} we have depicted Scenario II. As described
above this scenario again realizes
Gubser's proposal for the non-uniform black string branch that
starts at the Gregory-Laflamme point $(M,n)=(M_{\rm GL},1/(d-2))$.
But in this scenario the black hole branch continues into another
non-uniform string branch that continues for arbitrarily high masses
(following the proposal of \cite{Harmark:2002tr}).

\begin{figure}[ht]
\centerline{\epsfig{file=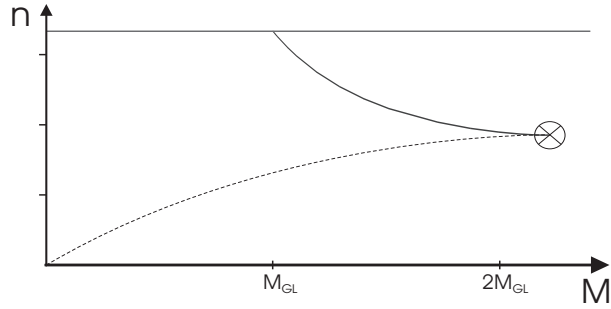,width=8cm,height=4cm}}
\caption{\small Scenario III. The scenario of Kol where the black
hole branch is connected to the non-uniform branch.}
\label{scenIII}
\end{figure}

In Figure \ref{scenIII} we have depicted Scenario III.
This is Kol's scenario where
the non-uniform string branch turns around
and meets the black hole branch. A smooth turn again seems
problematic, but in this case it seems natural to suppose that
there is a cusp precisely at the point where the phase transition
between the non-uniform string phase and the black hole phase occurs.%
\footnote{We thank Toby Wiseman for suggesting this to us.}

Evidence in support of this scenario was given in
Ref.~\cite{Kol:2003ja}, by numerically showing that
for sufficient non-uniformity the
non-uniform string branch has the property that the local geometry about the
minimal horizon sphere approaches the cone metric.
They argue furthermore that this implies that the non-uniform branch will
not continue beyond the endpoint of Wiseman's data at $M \simeq 2.3 M_{\rm GL}$. If true, this would imply that the other scenarios presented in this
section are not viable. 

\subsection{Other possible scenarios}

Each of the three scenarios described above
suffers from possible shortcomings. Kol's
scenario in Figure \ref{scenIII} does not have the needed Horowitz-Maeda
non-uniform string phase that Horowitz and Maeda argued for in
\cite{Horowitz:2001cz}. That can of course possibly turn into a virtue
in that the recent paper \cite{Choptuik:2003qd} failed to find an endpoint
of the classical decay, suggesting that a stable endpoint in fact
does not exists.

On the other hand, the scenarios I and II in Figure \ref{scenI}
and \ref{scenII} suffer from having a
seemingly unnatural U-turn on the curve.

\begin{figure}[ht]
\centerline{\epsfig{file=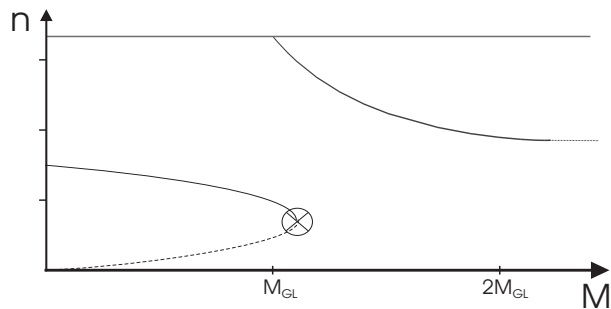,width=8cm,height=4cm}}
\caption{\small Scenario IV.}
\label{scenIV}
\end{figure}

One can instead imagine other scenarios which fit with the
data in Figure \ref{phase1}.
For example one can consider the new scenario depicted in
Figure \ref{scenIV}.
Note that this scenario is constructed so that it
has the Horowitz-Maeda non-uniform string branch and so that
the above-mentioned U-turn is absent.
Note also that the black hole branch
is disconnected from the non-uniform string branch originating
from the Gregory-Laflamme point $(M,n)=(M_{\rm GL},1/(d-2))$.

\begin{figure}[ht]
\centerline{\epsfig{file=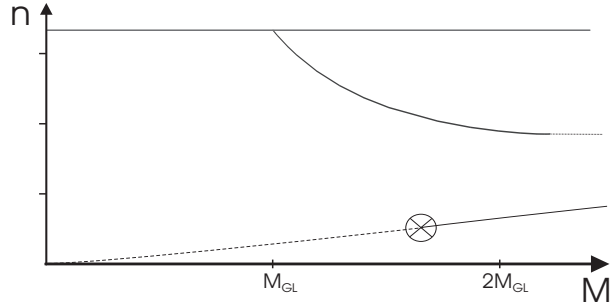,width=8cm,height=4cm}}
\caption{\small Scenario V.}
\label{scenV}
\end{figure}

\begin{figure}[ht]
\centerline{\epsfig{file=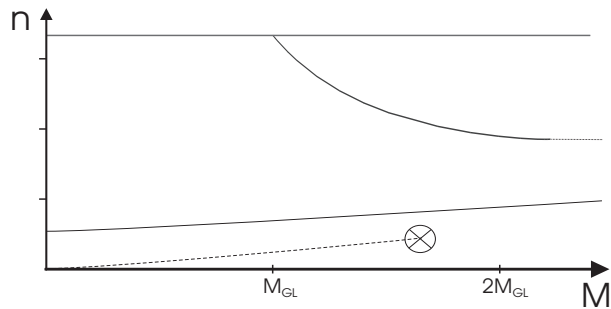,width=8cm,height=4cm}}
\caption{\small Scenario VI.}
\label{scenVI}
\end{figure}

Finally, we present the Scenarios V and VI in Figures
\ref{scenV} and \ref{scenVI}. These two scenarios
both have two non-uniform branches, each of which continues
for $M \rightarrow \infty$. Scenario V does not have the
conjectured Horowitz-Maeda strings while Scenario VI does.

\section{A remarkable near-linear behavior}
\label{secline}

\begin{figure}[ht]
\centerline{\epsfig{file=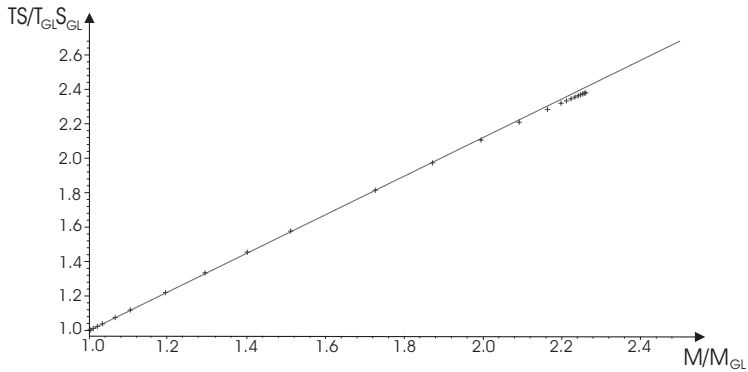,width=10cm,height=5cm}}
\caption{$TS$ versus $M$ plot for the non-uniform string branch.
The crosses are the data for Wiseman's solutions. The line is
a linear fit to the data.}
\label{plotmagic}
\end{figure}

In this section we report on a remarkable nearly linear behavior
of the non-uniform string branch, if one plots $TS$ versus
$M$. In Figure \ref{plotmagic} we have plotted Wiseman's data
in a $TS$ versus $M$ diagram, and we have fitted a line to the data points
in order to demonstrate how near the points falls to a line.

Figure \ref{plotmagic} shows that the $TS$ versus $M$ data falls very near
the line
\begin{equation}
\label{magicline}
\frac{TS}{T_{\rm GL} S_{\rm GL}} - 1
= x \left( \frac{M}{M_{\rm GL}} - 1 \right)
\ , \ \ x = 1.12 \ .
\end{equation}
Here $T_{\rm GL}$ and $S_{\rm GL}$ are
the temperature and entropy at the Gregory-Laflamme point
$(M,n)=(M_{\rm GL},1/(d-2))$.
Note that we have chosen the line so that it fits with the first half
of the set of data points.
Using now Smarr's formula \eqref{Smarra} with $d=5$
we see that \eqref{magicline} corresponds to the curve
\begin{equation}
\label{ncurve}
n = \frac{8}{3} (x-1) \frac{M_{\rm GL}}{M} + \frac{9-8x}{3} \ ,
\end{equation}
in the $(M,n)$ diagram. The explicit form of $n(M)$ above, enables us
to  integrate \eqref{intS} to find the entropy%
\footnote{More generally, we note that any branch represented by a line
in the $(M,TS)$ plot implies that $n(M) = \alpha + \beta/M $, which
in turn gives an entropy of the form $S(M)= S_0 (a M +b)^c$.}
\begin{equation}
S_{\rm w} = S_{\rm BS} (M_{\rm GL}) \left[ x \left( \frac{M}{M_{\rm GL}}
 -1\right) + 1  \right]^{\frac{3}{2x}} \ ,
\end{equation}
where the constant of integration is fixed by the intersection
point with the uniform string branch. The temperature then also follows
easily from Smarr's formula \eqref{Smarra}.

\begin{figure}[ht]
\centerline{\epsfig{file=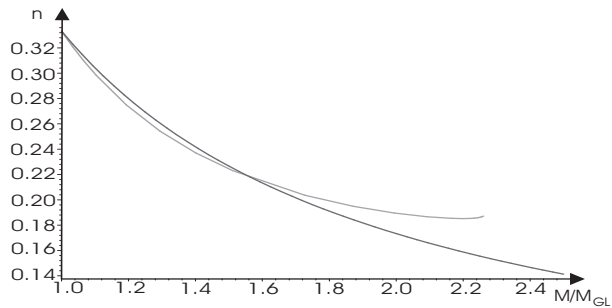,width=8cm,height=4cm}}
\caption{The non-uniform branch of Wiseman \cite{Wiseman:2002zc}
(dotted curve) compared
to the curve \eqref{ncurve} (solid curve).}
\label{plotcomp}
\end{figure}

We have plotted the curve \eqref{ncurve} along with Wiseman's data
in Figure \ref{plotcomp}. As one can see, Wiseman's data deviate
more and more from \eqref{ncurve} as the mass increases.
However, it is not clear whether Wiseman's
data is accurate enough to rule out that the exact solutions can
fall on a curve of the form \eqref{ncurve}.

Of course the above observations are of purely phenomenological
nature. However, it seems reasonable to expect that there
should be some theoretical reasoning
leading to the fact that the non-uniform string branch
have the linear behavior in Figure \ref{plotmagic},
at least to a good approximation.
This would be interesting to investigate further.

\section{Discussion and conclusions}
\label{secconcl}

In this paper, we have reported on several new results that
enhance the understanding of black holes and strings on cylinders.
This development builds partly on the new phase diagram and other
tools introduced in Ref.~\cite{Harmark:2003dg}.

We presented three main results.
Firstly, we proved that the metric
of any neutral and static black hole/string solution on a
cylinder can be written in the ansatz \eqref{ansatz},
an ansatz that was proposed in \cite{Harmark:2002tr}.
This proof is a generalization of the original
proof of Wiseman for $d=5$ in Ref.~\cite{Wiseman:2002ti}.
The ansatz \eqref{ansatz} is very simple since it
only depends on one function of two variables, which
means that most of the gauge freedom of the metric is
fixed.

Secondly, we showed that
the first law of thermodynamics
$\delta M = T \delta S$
holds for any neutral black hole/string on a cylinder.
This was proven using the ansatz \eqref{ansatz} and also
the Smarr formula \eqref{Smarra} of  \cite{Harmark:2003dg}.
That any curve of solutions obeys $\delta M = T \delta S$
implies that we do not expect a continuous
set of solutions between two given branches in the phase diagram.

The third result is the explicit construction of a
solution generator that takes any black/hole string solution
and transforms it to a new solution, thus giving a far richer
phase diagram.
This is a further development of an idea of Horowitz
\cite{Horowitz:2002dc} that a black hole/string on a cylinder
can be ``unwrapped'' to give new solutions.
We constructed the solution generator using the ansatz \eqref{ansatz}
which furthermore made it possible to write down explicit transformation
laws for all the thermodynamic quantities.

Beyond these new results, we have presented a comprehensive analysis
of the proposals for the phase structure in light of the new results
presented here and in Ref.~\cite{Harmark:2003dg}.
In addition to the three previously proposed scenarios, we
have listed three new scenarios for the phase structure, all of which
could conceivably be true.
We have also commented on the viability of the various
scenarios.

Finally, we presented a phenomenological observation:
Wiseman's numerical data \cite{Wiseman:2002zc}
for the non-uniform string branch
lie very near a straight line in a $TS$ versus $M$ diagram.
If the true non-uniform branch really
lies on a line, that would clearly suggest that only
Scenarios IV-VI would have a chance to be true (in principle
also Scenario III if the line terminates).
Moreover, it would present the tantalizing possibility
that perhaps {\sl all} branches of black holes/strings
on cylinders are lines in the $TS$ versus $M$
diagram. That would suggest that the black hole branch
lies on the curve given by $n=0$ in the $(M,n)$ phase diagram.

It will be highly interesting to uncover the final and correct
$(M,n)$ phase diagram for neutral black holes and strings on cylinders.
In the path of finding that, we are bound to learn more about several
important questions regarding black holes in General Relativity.
We can get a better understanding of the uniqueness properties of black
holes, i.e.
how many physically different solutions are there for a given
$M$ and $n$. We can learn about how ``physical'' a horizon really
is, i.e. does the black hole behave as a point particle, or
does it behave as an extended object when put on a cylinder?
Finally, we can learn much more about phase transitions
in General Relativity. Is Cosmic Censorship upheld or not for
a decaying light uniform string? Under which conditions
can the topology of an event horizon change?
We consider the ideas and results of Ref.~\cite{Harmark:2003dg} and this paper
as providing new guiding principles and
tools in the quest of understanding the
full phase structure of neutral black holes and strings on cylinders
so that all these important questions can
find their answer.

\section*{Acknowledgments}

We thank J. de Boer, H. Elvang, R. Emparan, G. Horowitz, V. Hubeny, M. Rangamani, S. Ross, K. Skenderis, M. Taylor, E. Verlinde and T. Wiseman for helpful
discussions.

\begin{appendix}

\section{Equivalence of boundary conditions for ans\"atze in $d=5$}
\label{appequi}

In this appendix, we give a refinement of the original proof
\cite{Wiseman:2002ti} that for $d=5$ the three-function conformal
ansatz \eqref{confans} of that paper
can be mapped onto the two-function ansatz \eqref{ansatz} proposed
in \cite{Harmark:2002tr}.

Our starting point is the conformal ansatz of \cite{Wiseman:2002ti}
\begin{equation}
\label{Wise}
ds^2 = -\frac{r^2}{m+ r^2} e^{2 A} dt^2 + e^{2B} (dr^2 + dz^2)
+ e^{2C} (m +r^2) d \Omega_3^2 \ ,
\end{equation}
where $A,B,C$ are functions of $(r,z)$, with asymptotic forms
\begin{equation}
\label{Wisebc}
A \simeq \frac{a_2}{r^2} \spa B \simeq \frac{b_2}{r^2} \spa C \simeq \frac{c_1}{r}
+ \frac{c_2}{r^2} \ .
\end{equation}
The new ingredient of the proof is that we first show that
by a suitable change of coordinates the leading term in the boundary
condition of the function $C$ can be eliminated, leaving
\begin{equation}
\label{asymC}
C \simeq \frac{c_2 + c_1^2/2}{r^2} \ ,
\end{equation}
while leaving the asymptotic forms of $A$ and $B$ unaffected.
The relevance of this step will become clear once we perform
the coordinate transformation to the two-function ansatz, as the
boundary condition \eqref{asymC} will be essential in order to
obtain  the correct boundary condition for the transformed metric.

To prove this statement we perform the following sequence of
coordinate transformations
\begin{equation}
\rho = \sqrt{m + r^2} \quad \rightarrow \quad
\hat \rho = \rho + c \quad \rightarrow \quad \hat \rho = \sqrt{m + \hat r^2} \ .
\end{equation}
After some algebra the resulting metric in $(\hat r,z)$ coordinates
becomes
\begin{equation}
\label{Wise2}
ds_6^2 = -\frac{\hat r^2}{m+ \hat r^2} e^{2\hat A} dt^2 +
e^{2\hat B}  [d\hat r^2 +  g^2( \hat r) dz^2]
+ e^{2 \hat C} (m +\hat r^2) d \Omega_3^2 \ ,
\end{equation}
with asymptotic conditions
\begin{equation}
\hat A \simeq \frac{a_2}{\hat r^2} \spa \hat B \simeq B \simeq \frac{b_2}{\hat r^2}
\spa
\hat C \simeq \frac{c_1 -c}{\hat r} + \frac{c_2 + c_1 (c_1-c) + c^2/2}{\hat r^2} \ ,
\end{equation}
and%
\footnote{The expressions for $\hat A$, $\hat C$ in terms of $A,C$ are easily
obtained but not important in the following.}
\begin{equation}
\label{factor}
g^2 (\hat r) \equiv \frac{e^{2  B} }{e^{2 \hat B}} =
\frac{ (1- \frac{m}{ \rho^2})}{(1- \frac{m}{ \hat \rho^2})}
=\frac{ (1- \frac{m}{ (\hat \rho-c)^2})}{(1- \frac{m}{ \hat \rho^2})}
= \frac{ (1- \frac{m}{ (\sqrt{m+ \hat r^2}-c)^2})}{(1- \frac{m}{ m + \hat r^2})}
\simeq 1 - \frac{2mc}{\hat r^3} \ .
\end{equation}
As a result we see that for the specific choice $c= c_1$ we can transform
to a metric where the boundary condition on $\hat C$ becomes
\begin{equation}
\hat{C} \simeq \frac{c_2 + c_1^2/2}{r^2} \ .
\end{equation}
However, in the process we have lost the conformal form
with one single function multiplying the factor $(d \hat r^2 + dz^2) $.

We therefore first need to transform back to the original conformal
form, which we achieve by changing variables
$\hat r = u(\tilde r)$
with the function $u$ satisfying the differential equation
 \begin{equation}
 \label{DE}
 \partial_{\tilde r} u (\tilde r)= g ( u(\tilde r)) \ .
 \end{equation}
 Then we find the form
\begin{equation}
\label{Wise3}
ds^2 = -\frac{u^2(\tilde r)}{m+ u^2(\tilde r)} e^{2\hat A (u(\tilde r))} dt^2 +
e^{2\hat B (u(\tilde r))}  [d\tilde r^2 +  dz^2 ]
+ e^{2 \hat C (u(\tilde r))} (m +u^2(\tilde r)) d \Omega_3^2 \ .
\end{equation}
The differential equation \eqref{DE} is pretty hard to solve,
but we only need to
examine it asymptotically in order to make sure that no $1/\tilde r$ terms
in the asymptotic conditions are induced. We thus consider its
asymptotic form
\begin{equation}
\partial_{\tilde r} u (\tilde r)= 1 - \frac{mc}{ u^3(\tilde r)} \ ,
\end{equation}
obtained using \eqref{factor}. The solution is
\begin{equation}
u(\tilde r) = \tilde r \left( 1 + \frac{  mc}{2 \tilde r^3} + \ldots  \right) \ ,
\end{equation}
and obeys  the desired property.

After this sequence of transformations leading to \eqref{Wise3}, we
change notation back to the symbols in \eqref{Wise} (but now we may
drop the $1/r$ term
in the asymptotic boundary condition of $C$). What remains is to
show that we can change coordinates to the two-function ansatz
\eqref{2fans}. We do not repeat these steps here, as they can
be obtained from the general treatment in Section \ref{secderiv}
by substituting the particular value $d=5$ in equations
\eqref{redef}-\eqref{delzh}. For that value, these general expressions
reduce to the steps that are equivalent to those
originally performed in \cite{Wiseman:2002ti}.

\section{A useful identity for static perturbations \label{iddM}}

In this appendix we derive an identity
which holds for static and Ricci-flat perturbations of a
static metric.
The identity
is the starting point in Section \ref{1stsec} for obtaining an important
relation on the variation of the mass $\delta M$
which is independent from the
generalized Smarr formula \eqref{Smarra}. The derivation and form
of this identity follows the corresponding one
in \cite{Wald:1984rg} (eq. (12.5.42)) where the four-dimensional case
was treated.

Consider in a $D$-dimensional space-time two
vector fields  $v$ and $w$ satisfying $D_\mu v^\mu = 0 $ and
$D_\mu w^\mu = 0$. Let $v$ and $w$ be commuting,
i.e. $\mathcal{L}_v w = \mathcal{L}_w v = 0$.
It is not difficult to show that as a consequence one has
$D_\mu v^{[\mu} w^{\nu]} = 0$
where $2 v^{[\mu} w^{\nu]} = v^\mu w^\nu - v^\nu w^\mu$.
Using Gauss theorem we then get that
\begin{equation}
\label{idS1}
\int_{S_1} dS_{\mu \nu} v^{[\mu} w^{\nu]} =
\int_{S_2} dS_{\mu \nu}v^{[\mu} w^{\nu]} \ .
\end{equation}

We now consider a static metric $g_{\mu \nu}$ and take $w$ as
the vector field $w = \frac{\partial}{\partial t}$
and let $v$ be given so that
$\frac{\partial}{\partial t} v^\mu = 0$.
First we see that $D_\mu w^\mu = 0$ since $\Gamma^\mu_{\mu 0} = 0$
because $g_{\mu \nu}$ is static with respect to $t=x^0$.
Secondly we see that
$(\mathcal{L}_w v)^\mu = \partial_0 v^\mu - v^\alpha \partial_\alpha w^\mu
= 0$ so $[v,w] = 0$.
Thus, we only need to require that $\partial_0 v^\mu = 0$ and
$D_\mu v^\mu = 0$ to get the desired properties of $v$ and $w$

Consider now a static
perturbation of the metric $\delta g_{\mu \nu}$
which leaves the Ricci tensor unchanged, i.e. $\delta R_{\mu \nu} = 0$.
Using the Palatini identity
$\delta R_{\mu \nu} =
D_\nu \delta  \Gamma^\alpha_{\mu\alpha}- D_\alpha \delta \Gamma^\alpha_{\mu\nu}$
we can write $g^{\mu \nu} \delta R_{\mu \nu}$ as
\begin{equation}
\label{vargR}
g^{\mu \nu} \delta R_{\mu \nu} = - D_\mu v^\mu
\spa
v^\mu = D_\nu \left[ g^{\mu \alpha} g^{\nu \beta} \delta g_{\alpha\beta}
- g^{\mu \nu} g^{\alpha \beta} \delta g_{\alpha\beta} \right] \ .
\end{equation}
Using now the requirement  $\delta R_{\mu \nu} = 0$ as well as
that $\delta g_{\mu \nu}$ is a static perturbation, we
find that the $v$ defined in \eqref{vargR} satisfies
 $\partial_0 v^\mu = 0$ and
$D_\mu v^\mu = 0$ as desired.

We are now ready to state the advertised identity which
generalizes Wald's identity (12.5.42) in \cite{Wald:1984rg}.
Consider a static metric $g_{\mu \nu}$ and a static perturbation
of the metric $\delta g_{\mu \nu}$ with $\delta R_{\mu \nu} = 0$.
Choose two oppositely oriented disjoint $(D-2)$-dimensional
surfaces $S_1$ and $S_2$ so that $\partial V = S_1 \cup S_2$ for
some $(D-1)$ dimensional volume $V$. Then we have that
\begin{equation}
\label{walds}
\int_{S_1} dS_{\mu \nu} v^\mu w^\nu  =
\int_{S_2} dS_{\mu \nu} v^\mu w^\nu \ ,
\end{equation}
with
\begin{equation}
\label{vwdef}
v^\mu = D_\nu \left[ g^{\mu \alpha} g^{\nu \beta} \delta g_{\alpha\beta}
- g^{\mu \nu} g^{\alpha \beta} \delta g_{\alpha\beta} \right]
\spa
w = \frac{\partial}{\partial t} \ .
\end{equation}

\end{appendix}

\addcontentsline{toc}{section}{References}


\begin{thebibliography}{10}

\bibitem{Harmark:2003dg}
T.~Harmark and N.~A. Obers, ``New phase diagram for black holes and strings on
  cylinders,''
\href{http://www.arXiv.org/abs/hep-th/0309116}{{\tt hep-th/0309116}}.
To appear in {\em Class. Quant. Grav.}. 

\bibitem{Kol:2003if}
B.~Kol, E.~Sorkin, and T.~Piran, ``Caged black holes: Black holes in
  compactified spacetimes {I} -- theory,''
\href{http://www.arXiv.org/abs/hep-th/0309190}{{\tt hep-th/0309190}}.

\bibitem{Gregory:1993vy}
R.~Gregory and R.~Laflamme, ``Black strings and p-branes are unstable,'' {\em
  Phys. Rev. Lett.} {\bf 70} (1993) 2837--2840,
\href{http://arXiv.org/abs/hep-th/9301052}{{\tt hep-th/9301052}}.

\bibitem{Gregory:1994bj}
R.~Gregory and R.~Laflamme, ``The instability of charged black strings and
  p-branes,'' {\em Nucl. Phys.} {\bf B428} (1994) 399--434,
\href{http://arXiv.org/abs/hep-th/9404071}{{\tt hep-th/9404071}}.

\bibitem{Horowitz:2001cz}
G.~T. Horowitz and K.~Maeda, ``Fate of the black string instability,'' {\em
  Phys. Rev. Lett.} {\bf 87} (2001) 131301,
\href{http://arXiv.org/abs/hep-th/0105111}{{\tt hep-th/0105111}}.

\bibitem{Choptuik:2003qd}
M.~W. Choptuik {\em et al.}, ``Towards the final fate of an unstable black
  string,'' {\em Phys. Rev.} {\bf D68} (2003) 044001,
\href{http://www.arXiv.org/abs/gr-qc/0304085}{{\tt gr-qc/0304085}}.

\bibitem{Gubser:2001ac}
S.~S. Gubser, ``On non-uniform black branes,'' {\em Class. Quant. Grav.} {\bf
  19} (2002) 4825--4844,
\href{http://www.arXiv.org/abs/hep-th/0110193}{{\tt hep-th/0110193}}.

\bibitem{Wiseman:2002zc}
T.~Wiseman, ``Static axisymmetric vacuum solutions and non-uniform black
  strings,'' {\em Class. Quant. Grav.} {\bf 20} (2003) 1137--1176,
\href{http://www.arXiv.org/abs/hep-th/0209051}{{\tt hep-th/0209051}}.

\bibitem{Gregory:1988nb}
R.~Gregory and R.~Laflamme, ``Hypercylindrical black holes,'' {\em Phys. Rev.}
  {\bf D37} (1988)
305.

\bibitem{Harmark:2002tr}
T.~Harmark and N.~A. Obers, ``Black holes on cylinders,'' {\em JHEP} {\bf 05}
  (2002) 032,
\href{http://www.arXiv.org/abs/hep-th/0204047}{{\tt hep-th/0204047}}.

\bibitem{Horowitz:2002dc}
G.~T. Horowitz, ``Playing with black strings,''
\href{http://www.arXiv.org/abs/hep-th/0205069}{{\tt hep-th/0205069}}.

\bibitem{Kol:2002xz}
B.~Kol, ``Topology change in general relativity and the black-hole black-string
  transition,''
\href{http://www.arXiv.org/abs/hep-th/0206220}{{\tt hep-th/0206220}}.

\bibitem{Wiseman:2002ti}
T.~Wiseman, ``From black strings to black holes,'' {\em Class. Quant. Grav.}
  {\bf 20} (2003) 1177--1186,
\href{http://www.arXiv.org/abs/hep-th/0211028}{{\tt hep-th/0211028}}.

\bibitem{Harmark:2003fz}
T.~Harmark and N.~A. Obers, ``Black holes and black strings on cylinders,''
  {\em Fortsch. Phys.} {\bf 51} (2003) 793--798,
\href{http://www.arXiv.org/abs/hep-th/0301020}{{\tt hep-th/0301020}}.

\bibitem{Kol:2003ja}
B.~Kol and T.~Wiseman, ``Evidence that highly non-uniform black strings have a
  conical waist,'' {\em Class. Quant. Grav.} {\bf 20} (2003) 3493--3504,
\href{http://www.arXiv.org/abs/hep-th/0304070}{{\tt hep-th/0304070}}.

\bibitem{Casadio:2000py}
R.~Casadio and B.~Harms, ``Black hole evaporation and large extra dimensions,''
  {\em Phys. Lett.} {\bf B487} (2000) 209--214,
\href{http://www.arXiv.org/abs/hep-th/0004004}{{\tt hep-th/0004004}}.

\bibitem{Casadio:2001dc}
R.~Casadio and B.~Harms, ``Black hole evaporation and compact extra
  dimensions,'' {\em Phys. Rev.} {\bf D64} (2001) 024016,
\href{http://www.arXiv.org/abs/hep-th/0101154}{{\tt hep-th/0101154}}.

\bibitem{Horowitz:2002ym}
G.~T. Horowitz and K.~Maeda, ``Inhomogeneous near-extremal black branes,'' {\em
  Phys. Rev.} {\bf D65} (2002) 104028,
\href{http://www.arXiv.org/abs/hep-th/0201241}{{\tt hep-th/0201241}}.

\bibitem{DeSmet:2002fv}
P.-J. De~Smet, ``Black holes on cylinders are not algebraically special,'' {\em
  Class. Quant. Grav.} {\bf 19} (2002) 4877--4896,
\href{http://www.arXiv.org/abs/hep-th/0206106}{{\tt hep-th/0206106}}.

\bibitem{Kol:2002dr}
B.~Kol, ``Speculative generalization of black hole uniqueness to higher
  dimensions,''
\href{http://www.arXiv.org/abs/hep-th/0208056}{{\tt hep-th/0208056}}.

\bibitem{Sorkin:2002nu}
E.~Sorkin and T.~Piran, ``Initial data for black holes and black strings in
  5d,'' {\em Phys. Rev. Lett.} {\bf 90} (2003) 171301,
\href{http://www.arXiv.org/abs/hep-th/0211210}{{\tt hep-th/0211210}}.

\bibitem{Frolov:2003kd}
A.~V. Frolov and V.~P. Frolov, ``Black holes in a compactified spacetime,''
  {\em Phys. Rev.} {\bf D67} (2003) 124025,
\href{http://www.arXiv.org/abs/hep-th/0302085}{{\tt hep-th/0302085}}.

\bibitem{Emparan:2003sy}
R.~Emparan and R.~C. Myers, ``Instability of ultra-spinning black holes,'' {\em
  JHEP} {\bf 09} (2003) 025,
\href{http://www.arXiv.org/abs/hep-th/0308056}{{\tt hep-th/0308056}}.

\bibitem{Traschen:2001pb}
J.~Traschen and D.~Fox, ``Tension perturbations of black brane spacetimes,''
\href{http://www.arXiv.org/abs/gr-qc/0103106}{{\tt gr-qc/0103106}}.

\bibitem{Townsend:2001rg}
P.~K. Townsend and M.~Zamaklar, ``The first law of black brane mechanics,''
  {\em Class. Quant. Grav.} {\bf 18} (2001) 5269--5286,
\href{http://www.arXiv.org/abs/hep-th/0107228}{{\tt hep-th/0107228}}.

\bibitem{Wald:1984rg}
R.~M. Wald, {\em General Relativity}.
\newblock The University of Chicago Press, 1984.

\end{thebibliography}

\providecommand{\href}[2]{#2}\begingroup\raggedright\endgroup

\end{document}